\documentclass[aps,prl,floats,floatfix,twocolumn,showpacs]{revtex4}
\usepackage{amsmath, amsfonts}
\usepackage{graphicx}
\usepackage{bm}

\newcommand{\td}[1]{{\rm d}#1}  

\newcommand{\SMetric}{g}     
\newcommand{\Lapse}{N}       
\newcommand{\Shift}{\beta}   

\newcommand{\ExCurv}{K}      
\newcommand{\TrExCurv}{K}    

\newcommand{\dtime}{\partial_t}   

\newcommand{\SFlatMetric}{f}      

\newcommand{\CF}{\psi}              

\newcommand{\CMetric}{{\tilde{g}}}    
\newcommand{\CLapse}{\tilde{N}}     
\newcommand{\dtCMetric}{\tilde{u}}  

\newcommand{\CA}{\tilde{A}}         
\newcommand{\CRicciS}{\tilde{R}}    

\newcommand{\CCD}{{\tilde\nabla}\!} 
\newcommand{\CCDu}{{\tilde\nabla}}  
\newcommand{\CLong}[1]{(\tilde{\mathbb L}{#1})} 

\newcommand{\Amp}{{\cal A}_{\rm max}}
\newcommand{\CAmp}{\tilde{{\cal A}}}
\newcommand{\CAmpFour}{\tilde{\cal A}_{c,4}}
\newcommand{\CAmpFive}{\tilde{\cal A}_{c,5}}
\newcommand{\maxCFc}{\CF_{crit}}

\newcommand{\Ch}{h}

\begin{document}

\title{Uniqueness and Non-uniqueness in the Einstein Constraints}

\author{Harald P. Pfeiffer${}^1$ and James W. York, Jr.${}^2$}

\affiliation{${}^1$Theoretical Astrophysics 130-33, California Institute of
Technology, Pasadena, CA 91125}

\affiliation{${}^2$Department of Physics, 
Cornell University, Ithaca, New York, 14853}

\date{\today}

\begin{abstract}
The conformal thin sandwich (CTS) equations are a set of four of the
Einstein equations, which generalize the Laplace-Poisson equation of
Newton's theory.  We examine numerically solutions of the CTS
equations describing perturbed Minkowski space, and find only one
solution.  However, we find {\em two} distinct solutions, one even
containing a black hole, when the lapse is determined by a fifth
elliptic equation through specification of the mean curvature.  While
the relationship of the two systems and their solutions is a
fundamental property of general relativity, this fairly simple example
of an elliptic system with non-unique solutions is also of broader
interest.
\end{abstract}


\maketitle

Within a space-plus-time
decomposition~\cite{Arnowitt-Deser-Misner:1962,York:1979}, Einstein's
equations, just as Maxwell's equations, split into evolution equations
and initial value equations. The latter constrain the initial data and
are generally solved by the use of potentials that satisfy elliptic
equations.  These potentials are expected to be unique, given suitable
boundary conditions.  For the Einstein constraints, indeed, in all
cases in which existence of solutions has been proved, the solution
was also shown to be unique.

However, we demonstrate in this paper that an important method of
solving the Einstein constraints through potentials, the {\em extended
conformal thin sandwich} (XCTS) equations~\cite{York:1999,
Pfeiffer-York:2003}, admits non-unique solutions.  Two solutions
appear to exist even for arbitrarily small perturbations away from
Minkowski space; one solution is just perturbed Minkowski space, but
the second one even contains a black hole for sufficiently {\em small}
perturbation.  The XCTS system thus provides a fairly simple example
of a nonlinear elliptic system with non-unique solutions, which may be
of interest outside general relativity.  Because the XCTS system lies
at the heart of most modern schemes to construct binary neutron
star~\cite{Wilson-Mathews:1989,Cook-Shapiro-Teukolsky:1996,Baumgarte-Cook-etal:1997,Marronetti-Mathews-Wilson:1998}
and binary black
hole~\cite{Gourgoulhon-Grandclement-Bonazzola:2001a,Grandclement-Gourgoulhon-Bonazzola:2001b,Cook:2002,Cook-Pfeiffer:2004}
initial data, our result is also relevant to the astrophysical problem
of computing inspiral waveforms for compact objects.

In general relativity, the initial data are the induced Riemannian
metric $\SMetric_{ij}$ and the extrinsic curvature $\ExCurv_{ij}$ 
on a spacelike hypersurface.
The data $(\SMetric_{ij},
\ExCurv_{ij})$ must satisfy four nonlinear constraint equations, 
and there are now two equivalent, complete approaches with which we can
solve the constraints.  One methods seeks $(\SMetric_{ij},
\ExCurv_{ij})$ directly~\cite{Pfeiffer-York:2003}, while the other,
the ``conformal thin sandwich'' (CTS) method~\cite{York:1999}, seeks
to construct $(\SMetric_{ij}, \dtime\SMetric_{ij})$.  In general
relativity, this difference is not trivial: $\ExCurv_{ij}$ depends
only on the embedding of the slice (see,
e.g. \cite{Choquet-Bruhat-York:1980}) while $\dtime\SMetric_{ij}$
depends also on the ambient coordinate neighborhood of the slice.
Both methods result in four coupled elliptic equations, and many
existence and uniqueness results are
known~\cite{Bartnik-Isenberg:grqc0405092}.

The {\em extended} conformal thin sandwich formalism adds a fifth
elliptic equation for the lapse-function, which arises from
specification of the time derivative of the mean curvature
$\TrExCurv\equiv \ExCurv_{ij}\SMetric^{ij}$.  In the extended system,
the free data consists entirely of ``variable \& velocity'' pairs $(q,
\dot q)$~\cite{Pfeiffer-York:2003}, as desirable for a Lagrangian
viewpoint, or a thin-sandwich viewpoint.  However, the fifth equation
couples strongly all five equations of the extended system,
complicating mathematical analysis, so that to our knowledge, there are no
rigorous mathematical results for the extended system, neither for
existence nor for uniqueness.  
Similar difficulties, as well as the non-uniqueness results reported in
this paper are not present for the standard initial value problem,
but arise when going beyond the mere initial value problem by placing
demands on the slicing in the ambient spacetime via specification of
$\dtime\TrExCurv$. 
Such issues connected with general relativity have
been called traditionally the ``problem of time.''

The CTS equations can be written as
\begin{align}
\label{eq:Ham2}
\CCDu^2\CF-\frac{1}{8}\CRicciS\CF-\frac{1}{12}\TrExCurv^2\CF^5 
+\frac{1}{8}\CF^{-7}\CA^{ij}\CA_{ij}&=0,\\
\label{eq:Mom2}
\CCD_j\Big(\frac{1}{2\CLapse}\CLong{\Shift}^{ij}\Big)
-\frac{2}{3}\CF^6\CCDu^i\TrExCurv
-\CCD_j\Big(\frac{1}{2\CLapse}\dtCMetric^{ij}\Big)
&=0.
\end{align}
Here, $\CCD$ and $\CRicciS$ are the covariant derivative and
the trace of the Ricci tensor associated with the {\em conformal
metric} $\CMetric_{ij}$, which is related to the physical metric by
$\SMetric_{ij}\!=\!\CF^{4}\CMetric_{ij}$.  Furthermore,
\begin{equation}
\CA^{ij}\equiv
\frac{1}{2\CLapse}
\Big(\CLong{\Shift}^{ij}-\dtCMetric^{ij}\Big)
=\CF^{10}\Big(\ExCurv^{ij}-\frac{1}{3}\SMetric^{ij}\TrExCurv\Big)
\end{equation}
represents the conformal trace-free extrinsic curvature,
$\CLong{\Shift}^{ij}\equiv 2\CCDu^{(i}\Shift^{j)}
-2/3\,\CMetric^{ij}\CCD_k\Shift^k$ is the conformal longitudinal
operator, and the traceless tensor
$\dtCMetric_{ij}\!\equiv\!\dtime\CMetric_{ij}$ represents the
time-derivative of the conformal metric.  Once the free data
$(\CMetric_{ij}, \dtCMetric_{ij}; \TrExCurv, \CLapse)$ are chosen,
Eqs.~(\ref{eq:Ham2}) and~(\ref{eq:Mom2}) are elliptic and determine
the conformal factor $\CF$ and the shift-vector $\Shift^i$.  The
conformal lapse $\CLapse$ is related to the physical lapse by
$\Lapse=\CF^6\CLapse$.

The elliptic equation for the lapse, which follows from the Einstein
evolution equation for $\TrExCurv$, can be written as
\begin{align}
\nonumber
\CCDu^2(\CLapse\CF^7)-(\CLapse\CF^7)\Big(\frac{1}{8}\CRicciS\!+\!\frac{5}{12}\TrExCurv^4\CF^4\!+\!\frac{7}{8}\CF^{-8}\CA^{ij}\CA_{ij}\Big)&=\\
\label{eq:Lapse}
-\CF^5(\dtime\TrExCurv-\Shift^k\partial_k\TrExCurv).&
\end{align}
Viewing $\dtime\TrExCurv$ instead of $\CLapse$ as freely specifiable,
the free data becomes $(\CMetric_{ij}, \dtCMetric_{ij}; \TrExCurv,
\partial_t\TrExCurv)$, and one must solve the coupled elliptic system
Eqs.~(\ref{eq:Ham2}), (\ref{eq:Mom2}) and~(\ref{eq:Lapse}).  This is
the XCTS system.

We will show non-uniqueness by explicitly constructing two solutions
for a certain choice of free data.  This choice is based on a
linearized quadrupolar gravitational wave~\cite{Teukolsky:1982} and
follows very closely~\cite{Pfeiffer-Kidder-etal:2005}.  We set
\begin{subequations}\label{eq:FreeData}
\begin{align}\label{eq:FreeData1}
\CMetric_{ij}&=\SFlatMetric_{ij}+\CAmp \left.\Ch_{ij}\right|_{t=0},\\
\dtCMetric_{ij}&=\CAmp\,\left.{\rm TF}_{\!\CMetric}\,\dtime \Ch_{ij}\right|_{t=0}, 
\\
\TrExCurv&=0,\\
\CLapse&=1,\quad\mbox{for CTS}\\
\label{eq:FreeData-5}
\dtime\TrExCurv&=0,\quad\mbox{for XCTS}
\end{align}
\end{subequations}
where $\SFlatMetric_{ij}$ represents the Euclidean metric, ${\rm
TF}_{\!\CMetric}$ means the trace-free part with respect to
$\CMetric_{ij}$, and
\begin{equation*}
\Ch_{ij}\td x^i\td x^j=-8r K \cos\Theta\sin^2\Theta\td r\td\phi
      -2r^2 L\sin^3\Theta \td\phi\td \Theta.
\end{equation*}
The functions $K$ and $L$ are are given by 
\begin{gather*}
K=r^{-2}\,F^{(2)}-3r^{-3}\,F^{(1)}+3r^{-4}\,F,\\
L=-r^{-1}\,F^{(3)}+2r^{-2}\,F^{(2)}-3r^{-3}\,F^{(1)}+3r^{-4}\,F,
\end{gather*}
where $\!F=\!F(r+t)\!=\!e^{-(r+t-r_0)^2/w^2}$ denotes the radial
profile, and $F^{(n)}\equiv \td^n F(x)/\td x^n$.  For small $\CAmp$,
this choice of $h_{ij}$ corresponds to a localized incoming
gravitational wave at radius $r_0\equiv 20$ with width $w\equiv 1$,
having odd parity and azimuthal quantum number $M=0$.

Being based on a solution to the linearized Einstein equations, the
free data (\ref{eq:FreeData}) satisfy the conformal thin sandwich
equations~(\ref{eq:Ham2}), (\ref{eq:Mom2}) and~(\ref{eq:Lapse}) to
linear order in $\CAmp$, but violate them to ${\cal O}(\CAmp^2)$.
Consequently, the XCTS equations have a solution which corrects the
free data by ${\cal O}(\CAmp^2)$, as demonstrated
in~\cite{Pfeiffer-Kidder-etal:2005}.
However, this is not the full story.

We employ the spectral elliptic solver described
in~\cite{Pfeiffer-Kidder-etal:2003} to solve
Eqs.~(\ref{eq:Ham2}),~(\ref{eq:Mom2}) and~(\ref{eq:Lapse}) in a
computational domain with outer radius $10^8$ with Dirichlet boundary
conditions, $\CF\!=\!\Lapse\!=\!1, \Shift^i\!=\!0$.  For smooth
problems, like the ones considered here, such a spectral method
results in exponential convergence, and allows construction of highly
accurate solutions.  All solutions presented in this paper converge
indeed exponentially; generally, the {\em maximum} residual of both
Hamiltonian and momentum constraint is $\lesssim 10^{-6}$, and other
quantities we mention below have converged to a similar level.  The
convergence rate diminishes along the ``upper branch'' of the extended
system at amplitudes $\CAmp\lesssim 0.2$, but the relative errors
remain $\lesssim\!10^{-4}$ even at $\CAmp=0.02$.  In all figures,
numerical errors are much smaller than the line-thickness, so that
essentially the ``true'' analytical solutions are plotted.

\begin{figure}
\includegraphics[scale=0.43]{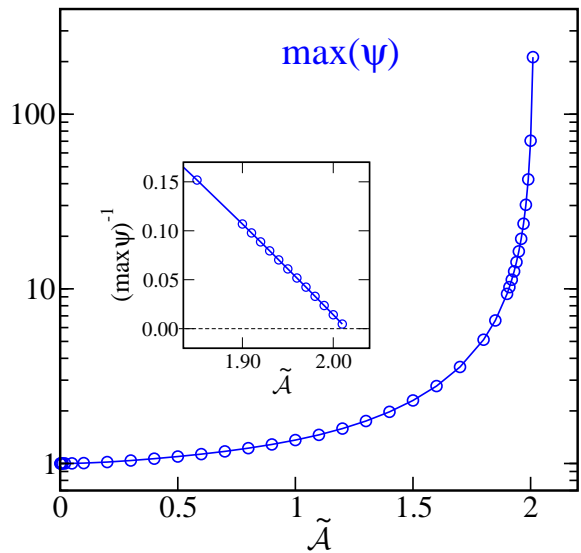}
\caption{\label{fig:4eqns}Solutions of the standard (four) conformal
thin sandwich equations vs. the conformal amplitude.  As the critical
amplitude $\CAmpFour\approx 2.01$ is approached, the conformal factor
grows to infinity.}
\end{figure}

Figure~\ref{fig:4eqns} presents solutions to the standard conformal thin
sandwich equations for various amplitudes $\CAmp$.  The conformal
factor $\CF$ diverges as $\CAmp$ approaches some critical amplitude.
For our particular choice $\CLapse\!=\!1$, $\TrExCurv\!=\!0$,
Eqs.~(\ref{eq:Ham2}) and (\ref{eq:Mom2}) reduce to a set of equations
which is very well understood mathematically~\cite{Cantor:1977,
Cantor-Brill:1981, Maxwell:2005}: Solutions are unique and exist iff
the Yamabe constant of $\CMetric_{ij}$ is positive.
Figure~\ref{fig:4eqns} suggests that the Yamabe constant of the metric
Eq.~(\ref{eq:FreeData1}) is positive for small $\CAmp$ and changes
sign at $\CAmp=\CAmpFour$, similar to an example given
in~\cite{Cantor-Brill:1981}.

\begin{figure}
\includegraphics[scale=0.43]{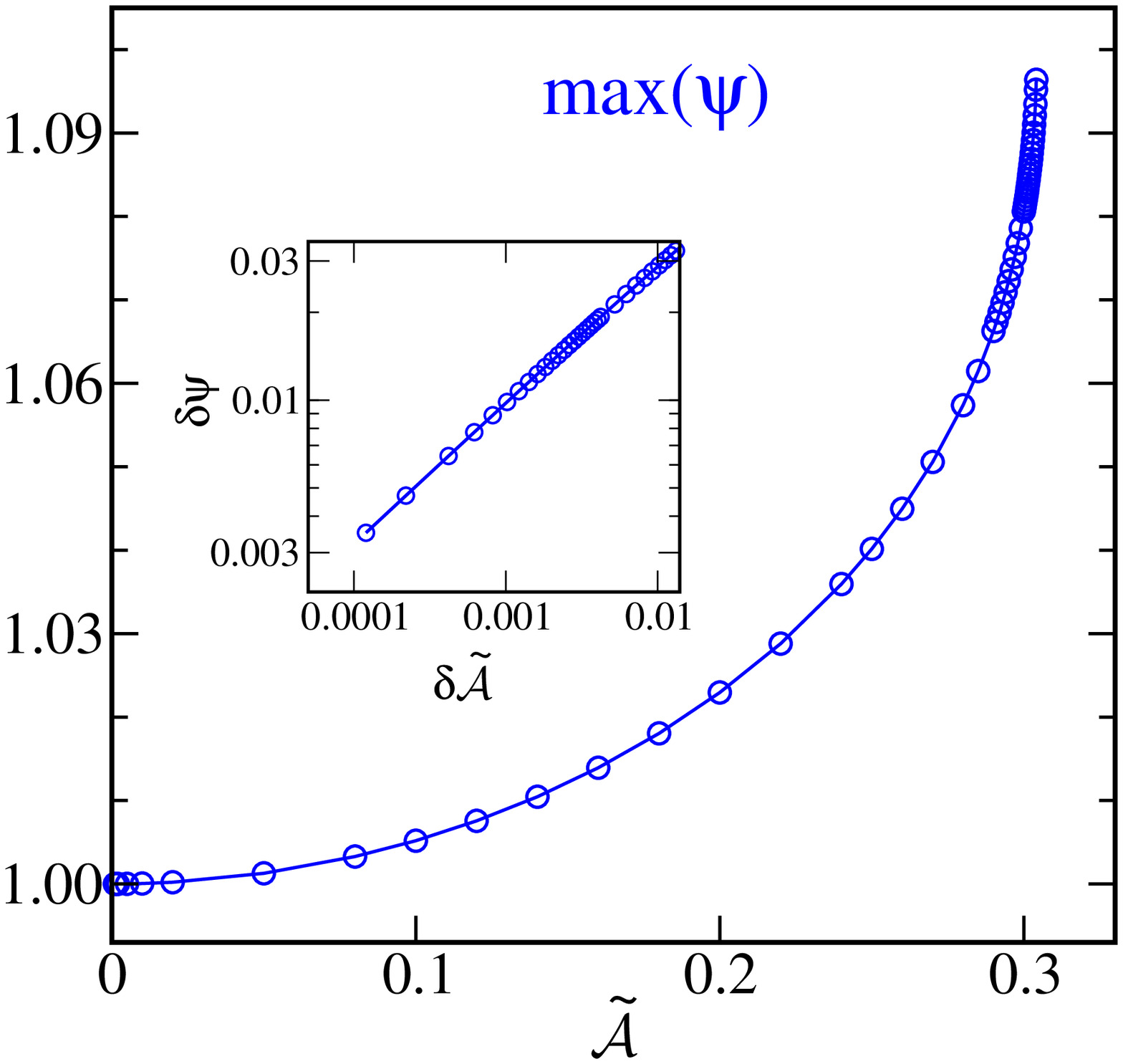}
\caption{\label{fig:5eqns}Solutions of the extended (five) conformal
thin sandwich equations vs. conformal amplitude.  The conformal factor
remains {\em finite} as the critical amplitude $\CAmpFive\approx0.304$
is approached, as confirmed by the inset ($\delta\CAmp=\CAmpFive\!-\!\CAmp$, $\delta\CF=\maxCFc-\max(\CF)$).}
\end{figure}

The situation is very different when solving the extended system.
Figure~\ref{fig:5eqns} presents the plot analogous to
Fig.~\ref{fig:4eqns} for this case.  As in the standard system, we
have not been able to find solutions beyond some critical amplitude.
However, here the similarity ends.  For the extended system, the
critical amplitude $\CAmpFive\approx 0.30422$ is much smaller than for
the standard system.  Therefore, existence of solutions cannot depend
simply on the sign of the Yamabe constant of $\CMetric_{ij}$.
Furthermore, as $\delta\CAmp\equiv\CAmpFive-\CAmp$ becomes small,
$\max\CF$ varies as $(\delta\CAmp)^{1/2}$ and approaches a {\em
finite} limit $\maxCFc\!\approx\! 1.0999$ (the numerical values for
$\CAmpFive$ and $\maxCFc$ were computed by fitting a parabola). The
log-log plot in the inset of Fig.~\ref{fig:5eqns} confirms these
statements.  Inspection of the solutions shows that as
$\delta\CAmp\!\to\!0$ {\em all} variables vary in proportion to
$(\delta\CAmp)^{1/2}$ at {\em any} spatial coordinate location $x$,
\begin{equation}
\label{eq:lowerbranch}
u_-(\CAmp; x)\approx u_{crit}(x)- v_{crit}(x)\sqrt{\delta\CAmp\,}.
\end{equation}
Here, $u_-$, $u_{crit}$ and $v_{crit}$ represent the vector of all
five variables $(\CF, \Shift^i, \CLapse)$, and the notation
indicates that $u_-$ depends on both $\CAmp$ and the spatial
coordinates $x$, whereas $u_{crit}$ and $v_{crit}$ are independent of
$\CAmp$ sufficiently close to $\CAmpFive$.

\begin{figure}
\includegraphics[scale=0.43]{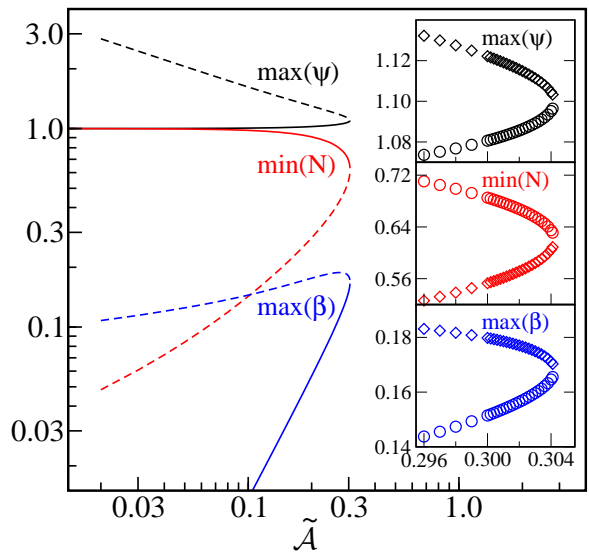}
\caption{\label{fig:5_vs_CA}Maximum of the conformal factor, minimum
of the physical lapse $\Lapse=\CF^6\CLapse$, and maximal magnitude of the shift
vs. conformal (unphysical) amplitude.  Solid/dashed curves represent
the lower/upper branch.  The insets show enlargements around the
critical point; circles/diamonds denote individual solutions along the
lower/upper branch.
}
\end{figure}

Because of the parabolic behavior close to $\CAmpFive$, we expect a
{\em second} branch of solutions, generalizing
Eq.~(\ref{eq:lowerbranch}) to
\begin{equation}\label{eq:bothbranches}
u_{\pm}(\CAmp;x)\approx u_{crit}(x) \pm
v_{crit}(x)\sqrt{\delta\CAmp\,}.
\end{equation}
The solutions found so far in Fig.~\ref{fig:5eqns} were obtained with
trivial initial guess $\CF\!=\!N=\!1$, $\Shift^i\!=\!0$, and
constitute the lower branch $u_{-}$ only; to converge to $u_+$, an
initial guess sufficiently close to $u_+$ is necessary.
Differentiation of Eq.~(\ref{eq:lowerbranch}) with respect to $\CAmp$
yields 
$v_{crit}=2(\delta\CAmp)^{1/2}\,\td u_-/\td\CAmp$,
so that,  sufficiently close to $\CAmpFive$,
\begin{equation}
\label{eq:upperbranch}
u_+(\CAmp; x)\approx u_-(\CAmp; x)
+ 4\,\delta\CAmp\;\frac{\,\td u_-(\CAmp; x)}{\td\CAmp}.
\end{equation}
Approximating $\td u_-/\td\CAmp$ by a finite-difference quotient of
solutions with $\CAmp\!=\!0.300$ and $\CAmp\!=\!0.301$, we compute the
right-hand-side of Eq.~(\ref{eq:upperbranch}) and use it as initial
guess for the elliptic solver at $\CAmp\!=\!0.300$.  The solver now
converges to a {\em different} solution; for example, $\max\CF\approx
1.12$.  This solution $u_+$ for one $\CAmp$ is then used as initial
guess for solutions with neighboring $\CAmp$ and repeating this
process, we move along the upper branch and compute solutions $u_+$
for a wide range of $\CAmp$.  Eventually, Fig.~\ref{fig:5_vs_CA}
emerges, which shows clearly the two distinct branches merging in a
parabola at $\CAmpFive$. 

\begin{figure}
\includegraphics[scale=0.43]{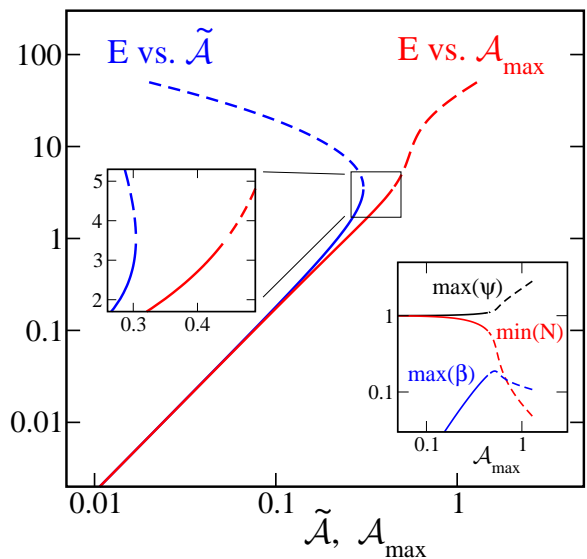}
\caption{\label{fig:5_vs_PhysA} Energy $E$ vs. unphysical
amplitude $\CAmp$ and physical amplitude $\Amp$.  The left inset
shows an enlargement around the critical point.  The right inset plots
the data of Fig.~\ref{fig:5_vs_CA} vs. $\Amp$.}
\end{figure}

The solutions $u_+$ are quite remarkable.  They exist over a wide
range of $\CAmp$, down to small $\CAmp$.  As $\CAmp$ decreases, $\CF$
increases everywhere; the increase is particularly pronounced close to
the maximum of $\CF$ which occurs in a ring in the equatorial plane at
radius close to $r_0$.  Consequently, with decreasing $\CAmp$, the
maxima of $\CF$ become increasingly pronounced and concentrated around
$r\approx r_0$ in the equatorial plane.  This necessitates higher
angular resolution in latitude and diminishes the convergence rate of
the spectral expansion, so that we choose to stop at $\CAmp\!=\!0.02$.
We note, however, that we do not see any fundamental indication that
the branch $u_+$ terminates; it appears that $u_+$ extents to
arbitrarily small $\CAmp$, with ever increasing $\CF$.  The physical
(ADM) energy $E$ of the computed initial data sets varies
approximately proportional to $\max\CF-1$ along $u_+$ (see also
Fig.~\ref{fig:5_vs_PhysA}).  At $\CAmp\!=\!0.02$, $E\!=\!50.012$ (in
units in which $r_0\!=\!20$).  This data set also contains an apparent
horizon with mass $M\!=\!\sqrt{\mbox{Area}/16\pi}\!=\!49.872$.  The
apparent horizon is oblate with equatorial radius $28.8$ and polar
radius $21.1$.  Apparent horizons are present for $\CAmp\lesssim
0.025$.

Figure~\ref{fig:5_vs_CA} shows clearly that the extended system
admits two solutions for the same free (conformal) data.  However, the
physical properties of the initial data sets depend on the {\em
physical} quantities $(\SMetric_{ij}, \ExCurv_{ij})$; for example, the
metric takes the form
\begin{equation}
\SMetric_{ij}=\CF^4\SFlatMetric_{ij}+\left(\CAmp\,\CF^4\right)\; \Ch_{ij}.
\end{equation}
This can be interpreted as the superposition of a gravitational wave
$\Ch_{ij}$ with {\em physical} amplitude $\CAmp\,\CF^4$ on a {\em
physical} background $\CF^4\SFlatMetric_{ij}$.  The conformal factor
$\CF$ is a function of space, and therefore $\CAmp\,\CF^4$ is also
spatially dependent.  To simplify, we take the maximum of $\CF$, which
coincides with the location of the gravitational wave at radius
$r_0\!=\!20$, and define a representative physical amplitude by
$\Amp\equiv\CAmp\,\max\CF^4$.

Figure~\ref{fig:5_vs_PhysA} plots energy $E$ both vs. $\CAmp$ and
$\Amp$; the qualitative difference between both definitions of
amplitude is apparent: The unphysical amplitude changes ``direction''
along the sequence of solutions leading to the largest energy at
smallest $\CAmp$.  In contrast, $\Amp$ is continuously {\em
increasing}, precisely one solution corresponds to each value of
$\Amp$, and large energies are obtained at large $\Amp$.  The family
of solutions passes smoothly through the critical point, and even for
arbitrarily large $\Amp$, a solution seems to exist. Thus, $\Amp$
represents the amplitude of the physical perturbation much more
faithfully than $\CAmp$.

We see that uniqueness of the problem is not a straightforward issue.
It depends on the question: for a given set of free data (this is, for
a given value of $\CAmp$), the elliptic equations may have {\em two}
solutions ($\CAmp<\CAmpFive$) or no solution at all
($\CAmp>\CAmpFive$).  However, for each value of the {\em physical}
amplitude $\Amp$ as described, precisely {\em one} solution exists
within the family of solutions considered here.

While the relationship between uniqueness and non-uniqueness of the
XCTS system as discussed above is intriguing, the major result of this
work lies in the fact that the XCTS system may have {\em two}
solutions for the same free data.  Besides the obvious importance of
this result to workers in mathematical relativity, we note that this
issue may complicate construction of astrophysical compact object
initial data, which relies on some form of the XCTS
equations~\cite{Wilson-Mathews:1989,Cook-Shapiro-Teukolsky:1996,Baumgarte-Cook-etal:1997,Marronetti-Mathews-Wilson:1998,Gourgoulhon-Grandclement-Bonazzola:2001a,Grandclement-Gourgoulhon-Bonazzola:2001b,Cook:2002,Cook-Pfeiffer:2004}.
Furthermore, the most common elliptic gauge
conditions~\cite{Smarr-York:1978} include precisely
Eq.~(\ref{eq:Lapse}) to determine the lapse-function.

The authors thank Lee Lindblom, Lawrence Kidder and Mark Scheel for
helpful discussions; the numerical code has been developed in
collaboration with Lawrence Kidder and Mark Scheel.  This work was
supported in part by NSF grants PHY-0244906 and PHY-0099568 to Caltech
and PHY-0407762, PHY-0311817 and PHY-0216986 to Cornell.  HP
gratefully acknowledges support through a fellowship of the Sherman
Fairchild Foundation.

\bibstyle{prl} 
\bibliography{Bibliography/References}

\begin{thebibliography}{21}
\expandafter\ifx\csname natexlab\endcsname\relax\def\natexlab#1{#1}\fi
\expandafter\ifx\csname bibnamefont\endcsname\relax
  \def\bibnamefont#1{#1}\fi
\expandafter\ifx\csname bibfnamefont\endcsname\relax
  \def\bibfnamefont#1{#1}\fi
\expandafter\ifx\csname citenamefont\endcsname\relax
  \def\citenamefont#1{#1}\fi
\expandafter\ifx\csname url\endcsname\relax
  \def\url#1{\texttt{#1}}\fi
\expandafter\ifx\csname urlprefix\endcsname\relax\def\urlprefix{URL }\fi
\providecommand{\bibinfo}[2]{#2}
\providecommand{\eprint}[2][]{\url{#2}}

\bibitem[{\citenamefont{Arnowitt et~al.}(1962)\citenamefont{Arnowitt, Deser,
  and Misner}}]{Arnowitt-Deser-Misner:1962}
\bibinfo{author}{\bibfnamefont{R.}~\bibnamefont{Arnowitt}},
  \bibinfo{author}{\bibfnamefont{S.}~\bibnamefont{Deser}}, \bibnamefont{and}
  \bibinfo{author}{\bibfnamefont{C.~W.} \bibnamefont{Misner}}, in
  \emph{\bibinfo{booktitle}{Gravitation}}, edited by
  \bibinfo{editor}{\bibfnamefont{L.}~\bibnamefont{Witten}}
  (\bibinfo{publisher}{Wiley, New York}, \bibinfo{year}{1962}).

\bibitem[{\citenamefont{York{,}~Jr.}(1979)}]{York:1979}
\bibinfo{author}{\bibfnamefont{J.~W.} \bibnamefont{York{,}~Jr.}}, in
  \emph{\bibinfo{booktitle}{Sources of Gravitational Radiation}}, edited by
  \bibinfo{editor}{\bibfnamefont{L.~L.} \bibnamefont{Smarr}}
  (\bibinfo{address}{Cambridge University Press, Cambridge, England},
  \bibinfo{year}{1979}), p.~\bibinfo{pages}{83}.

\bibitem[{\citenamefont{Pfeiffer and York{\, Jr.}}(2003)}]{Pfeiffer-York:2003}
\bibinfo{author}{\bibfnamefont{H.~P.} \bibnamefont{Pfeiffer}} \bibnamefont{and}
  \bibinfo{author}{\bibfnamefont{J.~W.} \bibnamefont{York{\, Jr.}}},
  \bibinfo{journal}{Phys. Rev. D} \textbf{\bibinfo{volume}{67}},
  \bibinfo{pages}{044022} (\bibinfo{year}{2003}).

\bibitem[{\citenamefont{York{, Jr.}}(1999)}]{York:1999}
\bibinfo{author}{\bibfnamefont{J.~W.} \bibnamefont{York{, Jr.}}},
  \bibinfo{journal}{Phys. Rev. Lett.} \textbf{\bibinfo{volume}{82}},
  \bibinfo{pages}{1350} (\bibinfo{year}{1999}).

\bibitem[{\citenamefont{Wilson and Mathews}(1989)}]{Wilson-Mathews:1989}
\bibinfo{author}{\bibfnamefont{J.~R.} \bibnamefont{Wilson}} \bibnamefont{and}
  \bibinfo{author}{\bibfnamefont{G.~J.} \bibnamefont{Mathews}}, in
  \emph{\bibinfo{booktitle}{Frontiers in Numerical Relativity}}, edited by
  \bibinfo{editor}{\bibfnamefont{C.~R.} \bibnamefont{Evans}},
  \bibinfo{editor}{\bibfnamefont{L.~S.} \bibnamefont{Finn}}, \bibnamefont{and}
  \bibinfo{editor}{\bibfnamefont{D.~W.} \bibnamefont{Hobill}}
  (\bibinfo{publisher}{Cambridge University Press},
  \bibinfo{address}{Cambridge, England, 1989}, \bibinfo{year}{1989}), pp.
  \bibinfo{pages}{306--314}.

\bibitem[{\citenamefont{Cook et~al.}(1996)\citenamefont{Cook, Shapiro, and
  Teukolsky}}]{Cook-Shapiro-Teukolsky:1996}
\bibinfo{author}{\bibfnamefont{G.~B.} \bibnamefont{Cook}},
  \bibinfo{author}{\bibfnamefont{S.~L.} \bibnamefont{Shapiro}},
  \bibnamefont{and} \bibinfo{author}{\bibfnamefont{S.~A.}
  \bibnamefont{Teukolsky}}, \bibinfo{journal}{Phys. Rev. D}
  \textbf{\bibinfo{volume}{53}}, \bibinfo{pages}{5533} (\bibinfo{year}{1996}).

\bibitem[{\citenamefont{Baumgarte et~al.}(1997)\citenamefont{Baumgarte, Cook,
  Scheel, Shapiro, and Teukolsky}}]{Baumgarte-Cook-etal:1997}
\bibinfo{author}{\bibfnamefont{T.~W.} \bibnamefont{Baumgarte}},
  \bibinfo{author}{\bibfnamefont{G.~B.} \bibnamefont{Cook}},
  \bibinfo{author}{\bibfnamefont{M.~A.} \bibnamefont{Scheel}},
  \bibinfo{author}{\bibfnamefont{S.~L.} \bibnamefont{Shapiro}},
  \bibnamefont{and} \bibinfo{author}{\bibfnamefont{S.~A.}
  \bibnamefont{Teukolsky}}, \bibinfo{journal}{Phys. Rev. Lett.}
  \textbf{\bibinfo{volume}{79}}, \bibinfo{pages}{1182} (\bibinfo{year}{1997}).

\bibitem[{\citenamefont{Marronetti et~al.}(1998)\citenamefont{Marronetti,
  Mathews, and Wilson}}]{Marronetti-Mathews-Wilson:1998}
\bibinfo{author}{\bibfnamefont{P.}~\bibnamefont{Marronetti}},
  \bibinfo{author}{\bibfnamefont{G.~J.} \bibnamefont{Mathews}},
  \bibnamefont{and} \bibinfo{author}{\bibfnamefont{J.~R.}
  \bibnamefont{Wilson}}, \bibinfo{journal}{Phys. Rev. D}
  \textbf{\bibinfo{volume}{58}}, \bibinfo{pages}{107503}
  (\bibinfo{year}{1998}).

\bibitem[{\citenamefont{Gourgoulhon et~al.}(2002)\citenamefont{Gourgoulhon,
  Grandcl{\'e}ment, and Bonazzola}}]{Gourgoulhon-Grandclement-Bonazzola:2001a}
\bibinfo{author}{\bibfnamefont{E.}~\bibnamefont{Gourgoulhon}},
  \bibinfo{author}{\bibfnamefont{P.}~\bibnamefont{Grandcl{\'e}ment}},
  \bibnamefont{and}
  \bibinfo{author}{\bibfnamefont{S.}~\bibnamefont{Bonazzola}},
  \bibinfo{journal}{Phys. Rev. D} \textbf{\bibinfo{volume}{65}},
  \bibinfo{pages}{044020} (\bibinfo{year}{2002}).

\bibitem[{\citenamefont{Cook}(2002)}]{Cook:2002}
\bibinfo{author}{\bibfnamefont{G.~B.} \bibnamefont{Cook}},
  \bibinfo{journal}{Phys. Rev. D} \textbf{\bibinfo{volume}{65}},
  \bibinfo{pages}{084003} (\bibinfo{year}{2002}).

\bibitem[{\citenamefont{Cook and Pfeiffer}(2004)}]{Cook-Pfeiffer:2004}
\bibinfo{author}{\bibfnamefont{G.~B.} \bibnamefont{Cook}} \bibnamefont{and}
  \bibinfo{author}{\bibfnamefont{H.~P.} \bibnamefont{Pfeiffer}},
  \bibinfo{journal}{Phys. Rev. D} \textbf{\bibinfo{volume}{70}},
  \bibinfo{pages}{104016} (\bibinfo{year}{2004}).

\bibitem[{\citenamefont{Grandcl{\'e}ment
  et~al.}(2002)\citenamefont{Grandcl{\'e}ment, Gourgoulhon, and
  Bonazzola}}]{Grandclement-Gourgoulhon-Bonazzola:2001b}
\bibinfo{author}{\bibfnamefont{P.}~\bibnamefont{Grandcl{\'e}ment}},
  \bibinfo{author}{\bibfnamefont{E.}~\bibnamefont{Gourgoulhon}},
  \bibnamefont{and}
  \bibinfo{author}{\bibfnamefont{S.}~\bibnamefont{Bonazzola}},
  \bibinfo{journal}{Phys. Rev. D} \textbf{\bibinfo{volume}{65}},
  \bibinfo{pages}{044021} (\bibinfo{year}{2002}).

\bibitem[{\citenamefont{Choquet-Bruhat and
  York}(1980)}]{Choquet-Bruhat-York:1980}
\bibinfo{author}{\bibfnamefont{Y.}~\bibnamefont{Choquet-Bruhat}}
  \bibnamefont{and} \bibinfo{author}{\bibfnamefont{J.~W.} \bibnamefont{York}},
  in \emph{\bibinfo{booktitle}{General Relativity and Gravitation: An
  {E}instein Centenary Survey}}, edited by
  \bibinfo{editor}{\bibfnamefont{A.}~\bibnamefont{Held}}
  (\bibinfo{publisher}{Plenum Press, New York}, \bibinfo{year}{1980}),
  vol.~\bibinfo{volume}{1}, pp. \bibinfo{pages}{99--172}.

\bibitem[{\citenamefont{Bartnik and
  Isenberg}(2004)}]{Bartnik-Isenberg:grqc0405092}
\bibinfo{author}{\bibfnamefont{R.}~\bibnamefont{Bartnik}} \bibnamefont{and}
  \bibinfo{author}{\bibfnamefont{J.}~\bibnamefont{Isenberg}},
  \bibinfo{journal}{gr-qc/0405092}  (\bibinfo{year}{2004}).

\bibitem[{\citenamefont{Teukolsky}(1982)}]{Teukolsky:1982}
\bibinfo{author}{\bibfnamefont{S.~A.} \bibnamefont{Teukolsky}},
  \bibinfo{journal}{Phys. Rev. D} \textbf{\bibinfo{volume}{26}},
  \bibinfo{pages}{745} (\bibinfo{year}{1982}).

\bibitem[{\citenamefont{Pfeiffer et~al.}(2005)\citenamefont{Pfeiffer, Kidder,
  Scheel, and Shoemaker}}]{Pfeiffer-Kidder-etal:2005}
\bibinfo{author}{\bibfnamefont{H.~P.} \bibnamefont{Pfeiffer}},
  \bibinfo{author}{\bibfnamefont{L.~E.} \bibnamefont{Kidder}},
  \bibinfo{author}{\bibfnamefont{M.~A.} \bibnamefont{Scheel}},
  \bibnamefont{and}
  \bibinfo{author}{\bibfnamefont{D.}~\bibnamefont{Shoemaker}},
  \bibinfo{journal}{Phys. Rev. D} \textbf{\bibinfo{volume}{71}},
  \bibinfo{pages}{024020} (\bibinfo{year}{2005}).

\bibitem[{\citenamefont{Pfeiffer et~al.}(2003)\citenamefont{Pfeiffer, Kidder,
  Scheel, and Teukolsky}}]{Pfeiffer-Kidder-etal:2003}
\bibinfo{author}{\bibfnamefont{H.~P.} \bibnamefont{Pfeiffer}},
  \bibinfo{author}{\bibfnamefont{L.~E.} \bibnamefont{Kidder}},
  \bibinfo{author}{\bibfnamefont{M.~A.} \bibnamefont{Scheel}},
  \bibnamefont{and} \bibinfo{author}{\bibfnamefont{S.~A.}
  \bibnamefont{Teukolsky}}, \bibinfo{journal}{Comput. Phys. Commun.}
  \textbf{\bibinfo{volume}{152}}, \bibinfo{pages}{253} (\bibinfo{year}{2003}).

\bibitem[{\citenamefont{Cantor}(1977)}]{Cantor:1977}
\bibinfo{author}{\bibfnamefont{M.}~\bibnamefont{Cantor}},
  \bibinfo{journal}{Comm. Math. Phys.} \textbf{\bibinfo{volume}{57}},
  \bibinfo{pages}{83} (\bibinfo{year}{1977}).

\bibitem[{\citenamefont{Maxwell}(2005)}]{Maxwell:2005}
\bibinfo{author}{\bibfnamefont{D.}~\bibnamefont{Maxwell}},
  \bibinfo{journal}{Comm. Math. Phys.} \textbf{\bibinfo{volume}{253}},
  \bibinfo{pages}{561} (\bibinfo{year}{2005}).

\bibitem[{\citenamefont{Cantor and Brill}(1981)}]{Cantor-Brill:1981}
\bibinfo{author}{\bibfnamefont{M.}~\bibnamefont{Cantor}} \bibnamefont{and}
  \bibinfo{author}{\bibfnamefont{D.}~\bibnamefont{Brill}},
  \bibinfo{journal}{Composit. Math.} \textbf{\bibinfo{volume}{43}},
  \bibinfo{pages}{317} (\bibinfo{year}{1981}).

\bibitem[{\citenamefont{Smarr and York{, Jr.}}(1978)}]{Smarr-York:1978}
\bibinfo{author}{\bibfnamefont{L.}~\bibnamefont{Smarr}} \bibnamefont{and}
  \bibinfo{author}{\bibfnamefont{J.~W.} \bibnamefont{York{, Jr.}}},
  \bibinfo{journal}{Phys. Rev. D} \textbf{\bibinfo{volume}{17}},
  \bibinfo{pages}{2529} (\bibinfo{year}{1978}).

\end{thebibliography}

\end{document}